# A Mechanism for Kaluza-Klein Number Violation in UED and Implications for LHC

**C.D. McMullen[a*] and S. Nandi[b†]**

[a] *Department of Physics, Louisiana School for Math, Science, and the Arts*
*Natchitoches, LA 71457, USA*

[b] *Department of Physics, Oklahoma State University*
*and Oklahoma Center for High Energy Physics,*
*Stillwater, OK 74078, USA*

**Abstract**

If the Higgs does not propagate into an otherwise universal extra dimension, then tree-level Kaluza-Klein number is not conserved in the Yukawa interactions. This leads to mixing between zero-mode fermions and their associated Kaluza-Klein excitations. The effect can be quite large for the top quark, since its mass (171.4 ± 2.1 GeV) is approximately one-half the current Tevatron mass bound (~350-400 GeV) for Kaluza-Klein excitations of quarks with one universal extra dimension. In contrast to the SM, the bi-unitary transformation that diagonalizes the mass matrix does not diagonalize the effective 4D Yukawa coupling matrix because of the presence of the 5D kinetic terms. Thus, Kaluza-Klein excitations of the fermions are not stable – e.g. they can decay to a zero-mode Higgs and zero-mode fermion. Furthermore, we find that this leads to violation of tree-level Kaluza-Klein number conservation in the gauge sector due to a unique feature of the 5D theory – namely, that the 5D quark multiplets consist of four-component vector-like quark fields. This allows Kaluza-Klein excitations of the fermions to be produced singly, and opens up new decay channels in the gauge sector. We compute the cross section for the production of the lowest-lying Kaluza-Klein excitations of the top and bottom quarks in the non-universal Higgs model at the Large Hadron Collider, and consider their subsequent decays. The effect is quite large: For example, pair production of Kaluza-Klein excitations of the bottom quark will lead to a pair of top quarks plus a pair of charged weak gauge bosons. Since the production is via the strong interaction, such final states stand out well against the Standard Model background, and are observable at the LHC for a wide range of compactification scales. Finally, we discuss the effects that such a model has on the *T*-parameter, and propose a model with maximal mixing where the compactification scale may be as large as a universal extra dimension.

[*] email: cmcmullen@lsmsa.edu
[†] email: s.nandi@okstate.edu

## 1. Introduction

In the original universal extra dimensions (UED) model [1], in which the Standard Model (SM) fields propagate into one large ($\mathrm{TeV}^{-1}$-size) extra dimension [2], tree-level Kaluza-Klein (KK) number is conserved. As a result, the KK excitations must be produced in pairs at high-energy colliders. Including radiative corrections to the masses of the KK excitations [3], the heavier KK excitations can decay to the lightest KK particle (LKP). However, they can not decay solely to SM particles.

We consider the possibility that the Higgs field does not propagate into one otherwise universal extra dimension (OUED). This can arise naturally if, for example, the Higgs simply propagates into a different extra dimension; but for generality, we also allow for the case that the Higgs is only partially excluded from the OUED. Such a non-universal Higgs (NUH) offers a way around the little hierarchy problem in the case of a UED bulk Higgs [4-5], in which the mass of the zero-mode Higgs field is pushed to the cut-off scale in 5D (or worse in higher dimensions).

One feature of the NUH model is that tree-level KK number is not conserved in the Yukawa interactions [6], which leads to mixing between zero-mode fermions and their associated KK excitations. Such inter-mode quark mixing is most significant for the third-generation charge 2/3 quark because the top quark mass (171.4 ± 2.1 GeV) [7] is roughly one-half the current Tevatron mass bound (~350-400 GeV) for KK excitations of quarks in the UED model [1,8-9]. The same bi-unitary transformation that diagonalizes the 5D mass matrix does not diagonalize the effective 4D Yukawa coupling matrix because the 5D kinetic terms for the KK masses is not present in the Yukawa coupling matrix. In consequence, the lowest-lying KK excitations of the top quark can decay at tree-level to a zero-mode top quark and a zero-mode Higgs.

Tree-level KK number violation in the Yukawa interactions leads to tree-level KK number violation in the gauge interactions in the NUH model. This is a unique feature of the 5D theory: Since the 5D quark multiplets consist of four-component vector-like quark fields, after orbifolding, for each left-chiral zero-mode quark field there are left-chiral as well as right chiral KK excitation that transforms as $SU(2)_L$ doublets in the effective 4D theory; and similarly for the right-chiral fields. Thus, there are inherently different couplings for the gauge interactions of the two towers of left-chiral KK excitations, and similarly for the two towers of right-chiral KK excitations. This gives rise to tree-level gauge interactions between zero-mode quarks and their associated KK excitations. These KK number-violating gauge interactions allow the KK excitations of the top quark to be produced singly at high-energy colliders, and provide new channels for their decays. For example, the lowest-lying KK excitations of the top quark can decay to a zero-mode top quark and zero-mode $Z$ boson or to a zero-mode bottom quark and a zero-mode $W$ boson.

A non-universal Higgs can have a significant effect on the $T$-parameter: Tree-level KK number non-conservation can lead to tree-level contributions to the $T$-parameter through the mixing between the zero-mode $W$ and $Z$, and their associated KK excitations, which can raise the bound on the compactification scale up to about 2 TeV [5]. However, there are two ways around this EW precision constraint. First, if the Higgs propagates partway into the OUED $(0 \le R_H \le R)$, the bound on $1/R$ from the $T$-parameter is more and more relaxed as $R_H \to R$. This follows since the mixing is zero in the limit $R_H \to R$ and maximal in the limit $R_H \to 0$. Secondly, it is possible to have maximal mixing $R_H \to 0$ yet avoid large tree-level contributions to the $T$-parameter with the following scenario: The fermions and gauge bosons – except for the Higgs – propagate into an OUED with a compactification scale $1/R_1$ as large as ~400 GeV; and the gauge bosons, including the Higgs – but not the fermions – propagate into a second, semi-universal extra dimension with a compactification scale $1/R_2$.

In this picture, $y_1 = 0$ for the Higgs and $0 \le y_1 \le R_1$ for the fermions and other gauge bosons, while $y_2 = 0$ for the fermions and $0 \le y_2 \le R_2$ for the gauge bosons and Higgs. The effective 4D Lagrangian density is obtained via a double integral over $y_1$ and $y_2$; a delta function $\delta(y_1)$ constrains the Higgs to lie at $y_1 = 0$ and a delta function $\delta(y_2)$ constrains the fermions to lie at $y_2 = 0$. The first delta function $\delta(y_1)$ provides tree-level KK number violation in the Yukawa interactions and subsequent features of maximal mixing in the NUH model, while the second delta function $\delta(y_2)$ provides tree-level KK number violation in the interactions between the fermions and gauge bosons. The latter leads to a strong bound on $1/R_2$ (~ 2-3 TeV), as in fermiphobic models where the gauge bosons propagate into the bulk while fermions are confined to the SM wall [10]. However, the bound on $1/R_1$ is analogous to bounds on the UED model (~400 GeV). The combination of constraints on $y_1$ and $y_2$, however, avoids large tree-level constraints on $1/R_1$ from the $T$-parameter. Since the $W^\pm$ propagates into two extra dimensions, its KK excitations need two indices, $W^{\pm(m,n)}$. The delta function $\delta(y_1)$, which would otherwise lead to tree-level mixing between the zero-mode $W^\pm$ and its KK excitations, instead leads to tree-level mixing between $W^{\pm(0,n)}$ and its KK excitations only; and no mixing between the zero-mode $W$ and $Z$, and their associated KK excitations.

Our paper is organized as follows. In Sec.'s 2-3, we develop the formalism for the Yukawa interactions and gauge interactions, respectively, between the zero-mode top quark and its lowest-lying KK excitations. In Sec. 4, we compute cross sections for the production of the lowest-lying KK excitations of the top and bottom quarks at the Large Hadron Collider (LHC), and calculate the branching ratios for their subsequent decays. In Sec. 5, we draw our conclusions.

## 2. Yukawa Interactions

We consider a model where the SM fermions and gauge bosons propagate into one $\text{TeV}^{-1}$-size extra dimension, but where the Higgs is fully or partially excluded from propagating into the OUED. We allow for both the simple case in which the Higgs either does not propagate into any extra dimensions or propagates into a different extra dimension, and for generality we also allow for the possibility that the Higgs may propagate partway into the OUED. The 5D quark multiplets consist of four-component vector-like quark fields, which can be decomposed into 4D two-component Weyl spinors via Fourier expansion about the OUED coordinate $y$:

$$Q_i(x^\mu, y) = \frac{1}{\sqrt{\pi R}}\left\{q_{iL}^{(0)}(x^\mu) + \sqrt{2}\sum_{n=1}^{\infty}\left[q_{iL}^{(n)}(x^\mu)\cos\left(\frac{ny}{R}\right) + q_{iR}^{(n)}(x^\mu)\sin\left(\frac{ny}{R}\right)\right]\right\}$$

$$U_i(x^\mu, y) = \frac{1}{\sqrt{\pi R}}\left\{u_{iR}^{(0)}(x^\mu) + \sqrt{2}\sum_{n=1}^{\infty}\left[u_{iR}^{(n)}(x^\mu)\cos\left(\frac{ny}{R}\right) + u_{iL}^{(n)}(x^\mu)\sin\left(\frac{ny}{R}\right)\right]\right\} \qquad (1)$$

$$D_i(x^\mu, y) = \frac{1}{\sqrt{\pi R}}\left\{d_{iR}^{(0)}(x^\mu) + \sqrt{2}\sum_{n=1}^{\infty}\left[d_{iR}^{(n)}(x^\mu)\cos\left(\frac{ny}{R}\right) + d_{iL}^{(n)}(x^\mu)\sin\left(\frac{ny}{R}\right)\right]\right\}$$

where $\{x^\mu\}$ are the usual 4D spacetime coordinates, the indices $\{i, j\} \in \{1,2,3\}$ represent the three quark generations, and $R$ is the radius of the OUED. There are two towers of KK excitations corresponding to each zero-mode quark: Associated with the left-chiral SM quark doublet $q_{iL}^{(0)}(x^\mu) \equiv \begin{pmatrix} u_i^{(0)}(x^\mu) \\ d_i^{(0)}(x^\mu) \end{pmatrix}_L$ are two KK quark doublets, a left-chiral doublet $q_{iL}^{(n)}(x^\mu)$ and a right-chiral doublet $q_{iR}^{(n)}(x^\mu)$; and associated with the right-chiral SM quark singlets $u_{iR}^{(0)}(x^\mu)$ and $d_{iR}^{(0)}(x^\mu)$ are KK singlets $u_{iL}^{(n)}(x^\mu)$, $u_{iR}^{(n)}(x^\mu)$, $d_{iL}^{(n)}(x^\mu)$, and $d_{iR}^{(n)}(x^\mu)$. Since there are not any zero-modes observed in the effective 4D theory corresponding to the right-chiral doublet $q_{iR}^{(n)}(x^\mu)$ and the left-chiral singlets $u_{iL}^{(n)}(x^\mu)$ and $d_{iL}^{(n)}(x^\mu)$, the extraneous zero-modes are projected out via the orbifold compactification choice $S^1/Z_2$ $(Z_2 : y \to -y)$.

Allowing for the most general non-universal Higgs (NUH), the Higgs may propagate partway into the OUED $(0 \le R_H < R)$, in which case the 5D Higgs doublet may be Fourier expanded about the OUED coordinate $y$ as

$$\Phi(x^\mu, y) = \frac{1}{\sqrt{\pi R_H}}\left[\Phi^{(0)}(x^\mu) + \sqrt{2}\sum_{n=1}^{\infty}\Phi^{(n)}(x^\mu)\cos\left(\frac{ny}{R_H}\right)\right] \qquad (2)$$

where the Higgs is constrained to lie within $0 \le y \le R_H$. The limit $R_H \to 0$ corresponds to the case where the Higgs either does not propagate into any extra dimensions or propagates into a different extra dimension. In order to obtain the zero-mode of the Higgs in the effective 4D theory, the 5D Higgs doublet must be even under the orbifold transformation $y \to -y$, i.e. $\Phi(x^\mu, -y) = \Phi(x^\mu, y)$. The zero-mode of the Higgs acquires the observed Higgs vev $\upsilon = \left(\sqrt{2}G_F\right)^{-1} = 247\,\text{GeV}$ [11], but any KK excitations of the Higgs are assumed not to acquire vev's.

The truncated mass matrix for the weak eigenstate zero-mode top quark and its lowest-lying KK excitations is [6]

$$
\left(\overline{q}_{3L}^{(0)} \quad \overline{q}_{3L}^{(1)} \quad \overline{u}_{3L}^{(1)}\right) D_n \begin{pmatrix} \frac{\widetilde{y}_4^{33}\upsilon}{\sqrt{2}} & 0 & \frac{\widetilde{y}_4^{33}\upsilon R}{\pi R_H}\sin\left(\frac{\pi R_H}{R}\right) \\ \frac{\widetilde{y}_4^{33}\upsilon R}{\pi R_H}\sin\left(\frac{\pi R_H}{R}\right) & \frac{1}{R} & \frac{\widetilde{y}_4^{33}\upsilon}{\sqrt{2}}\left[1+\frac{R}{2\pi R_H}\sin\left(\frac{2\pi R_H}{R}\right)\right] \\ 0 & \frac{\widetilde{y}_4^{33}\upsilon}{\sqrt{2}}\left[1-\frac{R}{2\pi R_H}\sin\left(\frac{2\pi R_H}{R}\right)\right] & -\frac{1}{R} \end{pmatrix} \begin{pmatrix} u_{3R}^{(0)} \\ q_{3R}^{(1)} \\ u_{3R}^{(1)} \end{pmatrix} \qquad (3)
$$

where $\widetilde{y}_4^{33}$ is the effective third-generation 4D Yukawa coupling and $\overline{q}_{3L}^{(0)}$, $\overline{q}_{3L}^{(1)}$, and $\overline{q}_{3R}^{(1)}$ represent the charge 2/3 components of their SU(2) doublets. The $(1,1)$ element provides the contribution of the Higgs vev to the zero-mode top quark, the $\pm 1/R$ terms are the kinetic terms for the KK excitations of the top quark, the $(1,2)$ and $(3,1)$ elements must be zero by $SU(2)$ invariance, and the other off-diagonal elements arise from tree-level violation of KK number conservation. There is a corresponding mass matrix for the Hermitian conjugate terms: $\left(\overline{u}_{3R}^{(0)}, \overline{q}_{3R}^{(1)}, \overline{u}_{3R}^{(1)}\right) M_t^{\dagger} \left(q_{3L}^{(0)}, q_{3L}^{(1)}, u_{3L}^{(1)}\right)^T$, where $M_t$ is the mass matrix in Eq. 3.

We incorporate a redefinition of the fields $\left(q_3^{(0)}, q_3^{(1)}, u_3^{(1)}\right)^T \to \left(q_3^{(0)}, q_3^{(1)}, \gamma_5 u_3^{(1)}\right)^T$ in order to accommodate the differing signs in the kinetic terms $(\pm 1/R)$; since these differing signs would lead to a negative eigenvalue, the role of the $\gamma_5$ is to ensure positive mass terms in the Lagrangian. The negative eigenvalue would otherwise lead to mass terms of the form $L_{mass}(x) = m_{u_3^{(0)}} \overline{u}_3^{(0)}(x) u_3^{(0)}(x) - m_{u_3^{(1)}} \overline{u}_3^{(1)}(x) u_3^{(1)}(x) + m_{q_3^{(1)}} \overline{q}_3^{(1)}(x) q_3^{(1)}(x)$. The redefinition of the KK field $u_3^{(1)}(x) \to \gamma_5 u_3^{(1)}(x)$ yields positive mass terms: $L_{mass}(x) \to m_{u_3^{(0)}} \overline{u}_3^{(0)}(x) u_3^{(0)}(x) + m_{u_3^{(1)}} \overline{u}_3^{(1)}(x) u_3^{(1)}(x) + m_{q_3^{(1)}} \overline{q}_3^{(1)}(x) q_3^{(1)}(x)$.

The $\gamma_5$'s can be absorbed into the left- and right-chiral fields as follows: $\gamma_5 u_{3L}^{(1)} = -u_{3L}^{(1)}$, $\gamma_5 u_{3R}^{(1)} = u_{3R}^{(1)}$, $-\overline{u}_{3L}^{(1)}\gamma_5 = -\overline{u}_{3L}^{(1)}$, and $-\overline{u}_{3R}^{(1)}\gamma_5 = \overline{u}_{3R}^{(1)}$. Thus, the redefinition $\left(q_{3L}^{(0)}, q_{3L}^{(1)}, u_{3L}^{(1)}\right)^T \to \left(q_{3L}^{(0)}, q_{3L}^{(1)}, \gamma_5 u_{3L}^{(1)}\right)^T$ can be written as $\left(q_{3L}^{(0)}, q_{3L}^{(1)}, u_{3L}^{(1)}\right)^T \to D_n \left(q_{3L}^{(0)}, q_{3L}^{(1)}, u_{3L}^{(1)}\right)^T$, where $D_n \equiv \mathrm{diag}\{1, 1, -1\}$, while the corresponding right-chiral fields $\left(q_{3R}^{(0)}, q_{3R}^{(1)}, u_{3R}^{(1)}\right)^T$ are unaffected by the transformation. Similarly, $\left(\overline{q}_{3L}^{(0)}, \overline{q}_{3L}^{(1)}, \overline{u}_{3L}^{(1)}\right) \to \left(\overline{q}_{3L}^{(0)}, \overline{q}_{3L}^{(1)}, \overline{u}_{3L}^{(1)}\right) D_n$, while $\left(\overline{q}_{3R}^{(0)}, \overline{q}_{3R}^{(1)}, \overline{u}_{3R}^{(1)}\right)$ experiences no sign changes.

The higher $(n \geq 2)$ KK excitations approximately decouple; mixing between the zero-mode top quark and its lowest-lying KK excitations is the dominant effect since the top quark mass (171.4 ± 2.1 GeV) is about one-half the current Tevatron mass bound (~350-400 GeV) for KK excitations of quarks propagating into one UED. Since first and second generation zero-modes are observed to be much lighter, we neglect mixing with the lighter generations; similarly, we also neglect mixing in the bottom sector.

The limit $R_H \to 0$ corresponds to the simple, natural case where the Higgs either does not propagate into any extra dimensions or propagates into a different extra dimension than the other SM fields. Mixing between the zero-mode top quark and its lowest-lying KK excitations is maximal in this limit, for which the truncated mixing matrix for the zero-mode top quark and its first KK excitations simplifies to

$$\begin{pmatrix} \bar{q}_{3L}^{(0)} & \bar{q}_{3L}^{(1)} & \bar{u}_{3L}^{(1)} \end{pmatrix} D_n \begin{pmatrix} \dfrac{\tilde{y}_4^{33}\upsilon}{\sqrt{2}} & 0 & \tilde{y}_4^{33}\upsilon \\ \tilde{y}_4^{33}\upsilon & \dfrac{1}{R} & \tilde{y}_4^{33}\upsilon\sqrt{2} \\ 0 & 0 & -\dfrac{1}{R} \end{pmatrix} \begin{pmatrix} u_{3R}^{(0)} \\ q_{3R}^{(1)} \\ u_{3R}^{(1)} \end{pmatrix} \tag{4}$$

In the other extreme, $R_H \to R$, the SM quarks decouple from their associated KK excitations. In this limit there is only mixing between KK excitations of the same mode, and the tree-level masses of the KK fields of the same mode become degenerate.

As a result of the off-diagonal terms in the mass matrix (Eq. 3), the zero-mode of the top quark is not the observed top quark; rather, the observed top quark, with a mass of 171.4 ± 2.1 GeV, is actually a linear combination of the zero-mode top quark and its lowest-lying KK excitations. We denote the observed mass eigenstates as follows: $t$ represents the observed top quark, with a mass of 171.4 ± 2.1 GeV, while $t_1^\bullet$ and $t_1^\circ$ represent the mass eigenstates for the lowest-lying KK excitations of the top quark. The $t_1^\bullet$ and $t_1^\circ$ do not couple to the zero-modes in quite the same way; even in their interactions with gluons, in the limit that the top quark is massless, the couplings to the $t_1^\bullet$ and $t_1^\circ$ differ via a projection operator $(1 \mp \gamma_5)/2$ [8].

These mass eigenstates $\{t, t_1^\bullet, t_1^\circ\}$ are determined from the weak eigenstates $\{q_3^{(0)}, q_3^{(1)}, u_3^{(1)}\}$ via a bi-unitary transformation:

$$\begin{aligned} \left(t, t_1^\bullet, t_1^\circ\right)_L^T &= U_L^{-1}\left(q_3^{(0)}, q_3^{(1)}, u_3^{(1)}\right)_L^T \\ \left(t, t_1^\bullet, t_1^\circ\right)_R^T &= U_R^{-1}\left(u_3^{(0)}, q_3^{(1)}, u_3^{(1)}\right)_R^T \end{aligned} \tag{5}$$

The matrix for the effective 4D Yukawa interactions $Y$ in the weak eigenstate basis is obtained from $\upsilon Y/\sqrt{2} = M_t - K$ where

$$K \equiv \begin{pmatrix} 0 & 0 & 0 \\ 0 & 1/R & 0 \\ 0 & 0 & -1/R \end{pmatrix} \tag{6}$$

Compared to the mass matrix, the matrix for the effective 4D Yukawa interactions does not have $\pm 1/R$ terms in the $(2,2)$ and $(3,3)$ elements that come from the 5D kinetic terms. Thus, one unique feature of the NUH, as compared to quark mixing in the SM, is that the bi-unitary transformation that diagonalizes the 5D mass matrix does not also diagonalize the corresponding effective 4D Yukawa coupling matrix. This leads to off-diagonal Yukawa interactions between the top quark, its lowest-lying KK excitations, and the Higgs boson.

The effective 4D matrix for the Yukawa couplings in the mass eigenstate basis is related to $Y$ through the substitutions in Eq. 5:

$$\begin{aligned} \left(\bar{t}_L,\bar{t}^{\bullet}_{1L},\bar{t}^{\circ}_{1L}\right)G^{Y}_{LR}\left(t_R,t^{\bullet}_{1R},t^{\circ}_{1R}\right)^T &= \left(\bar{q}^{(0)}_{3L},\bar{q}^{(1)}_{3L},\bar{u}^{(1)}_{3L}\right)D_n Y\left(u^{(0)}_{3R},q^{(1)}_{3R},u^{(1)}_{3R}\right)^T \\ \left(\bar{t}_R,\bar{t}^{\bullet}_{1R},\bar{t}^{\circ}_{1R}\right)G^{Y}_{RL}\left(t_L,t^{\bullet}_{1L},t^{\circ}_{1L}\right)^T &= \left(\bar{u}^{(0)}_{3R},\bar{q}^{(1)}_{3R},\bar{u}^{(1)}_{3R}\right)Y^{\dagger}D_n\left(q^{(0)}_{3L},q^{(1)}_{3L},u^{(1)}_{3L}\right)^T \end{aligned} \tag{7}$$

The matrices $G^{Y}_{LR}$ and $G^{Y}_{RL}$ equal $D_n U_L^{-1} Y U_R$ and $U_R^{-1}Y^{\dagger}U_L D_n$, respectively, where the diagonal matrix $D_n$ accounts for the effects from the redefinition of the fields $\left(q^{(0)}_3,q^{(1)}_3,u^{(1)}_3\right)^T \to \left(q^{(0)}_3,q^{(1)}_3,\gamma_5 u^{(1)}_3\right)^T$; $D_n$ affects the signs of the $(1,3)$, $(2,3)$, $(3,1)$, $(3,2)$ and (3,3) elements. The left- and right-chiral fields combine to form the effective 4D couplings of the mass eigenstates $\{t,t^{\bullet}_1,t^{\circ}_1\}$ to the Higgs boson:

$$\left(\bar{t},\bar{t}^{\bullet}_1,\bar{t}^{\circ}_1\right)G^{H}\left(t,t^{\bullet}_1,t^{\circ}_1\right)^T = \left(\bar{t}_L,\bar{t}^{\bullet}_{1L},\bar{t}^{\circ}_{1L}\right)G^{Y}_{LR}\left(t_R,t^{\bullet}_{1R},t^{\circ}_{1R}\right)^T + \left(\bar{t}_R,\bar{t}^{\bullet}_{1R},\bar{t}^{\circ}_{1R}\right)G^{Y}_{RL}\left(t_L,t^{\bullet}_{1L},t^{\circ}_{1L}\right)^T \tag{8}$$

In general, the matrix $G^H$ includes scalar couplings $\left(G^{Y}_{LR}+G^{Y}_{RL}\right)/2$ as well as pseudoscalar couplings $\left(G^{Y}_{LR}-G^{Y}_{RL}\right)\gamma_5/2$. As it turns out, for the natural case $R_H=0$, the pseudoscalar couplings are identically zero (as shown in the Appendix). For $0<R_H<1$, the pseudoscalar couplings could lead to CP violation.

We present a numerical example to illustrate this bi-unitary transformation, choosing $R=(500\,\mathrm{GeV})^{-1}$, $\tilde{y}^{33}_4=1.35$, and $R_H=0$. The $3\times 3$ top quark mass matrix is

$$M_t\left(R_H=0, 1/R=500\,\mathrm{GeV}, \tilde{y}^{33}_4=1.35\right)=\begin{pmatrix} 236 & 0 & 333 \\ 333 & 500 & 472 \\ 0 & 0 & -500 \end{pmatrix}\mathrm{GeV} \tag{9}$$

We diagonalize $M_t M_t^{\dagger}$ to find the corresponding mass-squared eigenvalues by brute force solution to the characteristic cubic: $m_t^2=(172\,\mathrm{GeV})^2$, $m^2_{t^{\bullet}_1}=(909\,\mathrm{GeV})^2$, and $m^2_{t^{\circ}_1}=(376\,\mathrm{GeV})^2$. We compute the eigenvectors of $M_t M_t^{\dagger}$ and $M_t^{\dagger}M_t$, respectively, to construct the unitary matrices, $U_L$ and $U_R$:

$$U_L=\begin{pmatrix} 0.864 & 0.397 & 0.311 \\ -0.164 & 0.804 & -0.571 \\ 0.477 & -0.442 & -0.760 \end{pmatrix} \quad , \quad U_R=\begin{pmatrix} 0.864 & 0.397 & 0.311 \\ -0.477 & 0.442 & 0.760 \\ -0.164 & 0.804 & -0.571 \end{pmatrix} \tag{10}$$

These unitary matrices correspond to the basis in Eq. 4.[‡] The diagonalized matrix is

$$D_n U_L^{-1} M_t U_R = \begin{pmatrix} 172 & 0 & 0 \\ 0 & 909 & 0 \\ 0 & 0 & 376 \end{pmatrix} \text{GeV} \tag{11}$$

The mass terms are all positive, in accordance with our redefinition of the fields $\left(q_{3L}^{(0)}, q_{3L}^{(1)}, u_{3L}^{(1)}\right)^T \to D_n \left(q_{3L}^{(0)}, q_{3L}^{(1)}, u_{3L}^{(1)}\right)^T$. The matrix for the effective 4D Yukawa couplings of the mass eigenstates $\left\{t, t_1^\bullet, t_1^\circ\right\}$ to the Higgs boson is

$$G^H\left(R_H = 0, 1/R = 500\,\text{GeV}, \tilde{y}_4^{33} = 1.35\right) = \begin{pmatrix} 0.539 & 1.31 & -0.424 \\ 1.31 & 3.18 & -1.03 \\ -0.424 & -1.03 & 0.333 \end{pmatrix} \tag{12}$$

As expected, only the scalar couplings to the Higgs survive; all of the pseudoscalar couplings to the Higgs are numerically zero.

The nonzero couplings of the mass eigenstates $\left\{t, t_1^\bullet, t_1^\circ\right\}$ to the Higgs boson are illustrated in Fig. 1 as a function of the compactification scale $1/R$ with $R_H = 0$. In this and forthfollowing plots, we use the values of $\tilde{y}_4^{33}$ that correspond to the central allowed region – found by demanding that the lightest mass eigenvalue correspond to the observed top quark mass – of Fig.'s 1-2 of Ref. [6]. The $ttH$ coupling (black rectangles ■), corresponding to $G_{11}^H$, differs from its SM value, especially for smaller values of $1/R$. The $tt_1^\bullet H$ coupling (gray rectangles ■), corresponding to $G_{12}^H$, is virtually constant in this range of parameter-space, whereas the $tt_1^\circ H$ coupling varies noticeably.

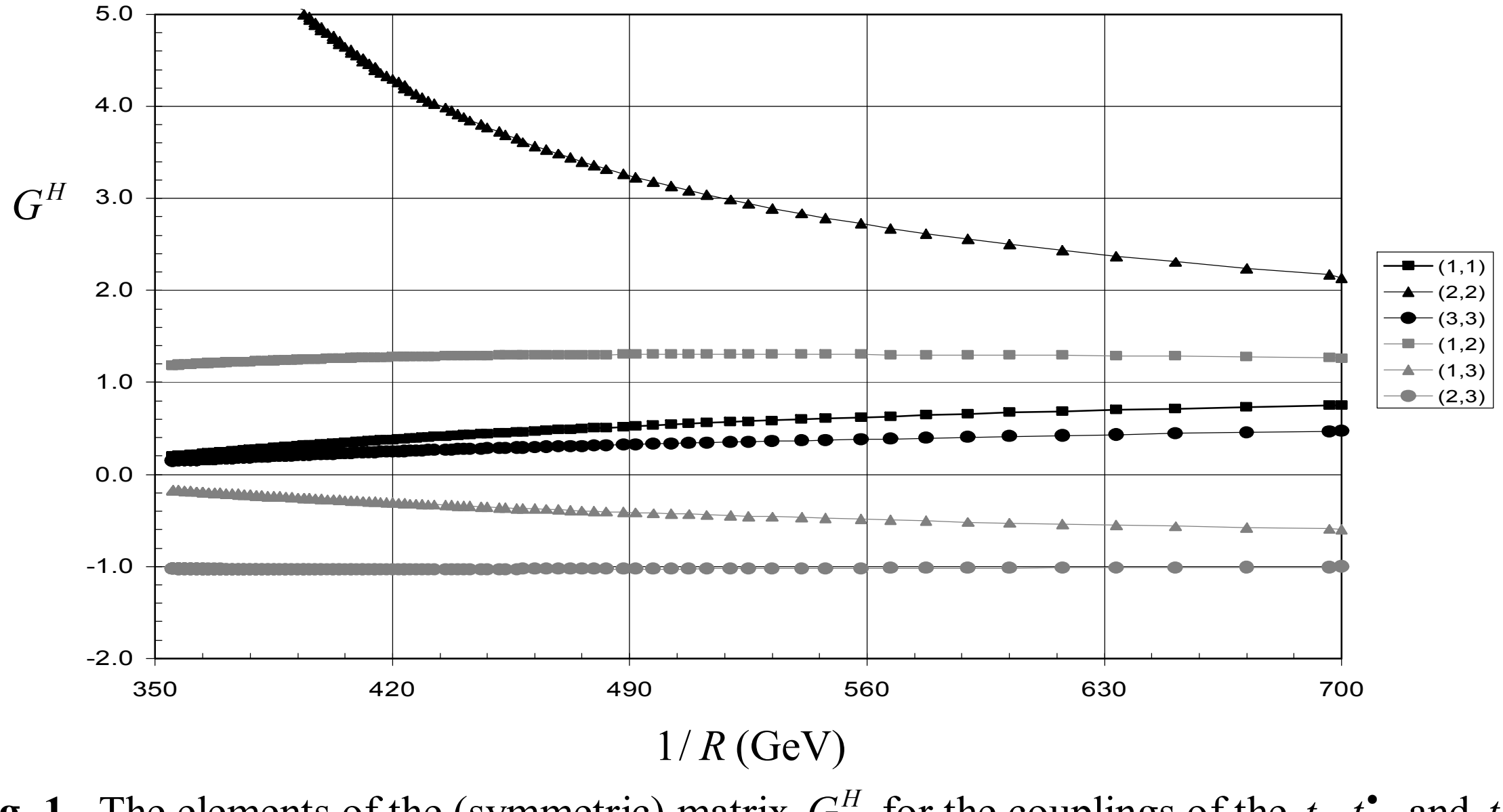

**Fig. 1**. The elements of the (symmetric) matrix $G^H$ for the couplings of the $t$, $t_1^\bullet$, and $t_1^\circ$ to the Higgs are plotted as a function of the compactification scale $1/R$ in the NUH model with $R_H = 0$ for the central allowed values of $\tilde{y}_4^{33}$ in the basis $\left(t, t_1^\bullet, t_1^\circ\right)$. (Only the scalar couplings survive; the pseudoscalar couplings all equal zero.)

[‡] Comparing with Eq. 16 in Ref. [6] (where only the eigenvalues were used in calculations) the eigenvectors had not been reordered to match this choice of basis.

### 3. Gauge Interactions

The effective 4D Lagrangian density for the interactions of the charge 2/3, third-generation quark fields with the electroweak gauge fields is obtained by integrating the corresponding 5D Lagrangian density over the OUED:

$$
\begin{aligned}
L^G\left(x^\mu\right) = & \, i\int_{y=0}^{\pi R} \overline{Q}_{5L}\left(x^\mu,y\right)\Gamma^M\left[\partial_M + i\frac{g\tau^a}{2}W_M^a\left(x^\mu,y\right)+i\frac{g'Y_L}{2}B_M\left(x^\mu,y\right)\right]Q_{5L}\left(x^\mu,y\right)dy \\
& + i\int_{y=0}^{\pi R} \overline{U}_{5L}\left(x^\mu,y\right)\Gamma^M\left[\partial_M + i\frac{g'Y_R}{2}B_M\left(x^\mu,y\right)\right]U_{5L}\left(x^\mu,y\right)dy \\
& + i\int_{y=0}^{\pi R} \overline{Q}_{5R}\left(x^\mu,y\right)\Gamma^M\left[\partial_M + i\frac{g\tau^a}{2}W_M^a\left(x^\mu,y\right)+i\frac{g'Y_L}{2}B_M\left(x^\mu,y\right)\right]Q_{5R}\left(x^\mu,y\right)dy \\
& + i\int_{y=0}^{\pi R} \overline{U}_{5R}\left(x^\mu,y\right)\Gamma^M\left[\partial_M + i\frac{g'Y_R}{2}B_M\left(x^\mu,y\right)\right]U_{5R}\left(x^\mu,y\right)dy
\end{aligned}
\tag{13}
$$

where lower-case Greek indices $\{\mu,\nu\}\in\{0,1,2,3\}$ are the usual 4D spacetime indices, upper-case Latin indices $\{M,N\}\in\{0,1,2,3,4\}$ are 5D spacetime indices, and $\{\Gamma^M\}$ are the 5D generalization of the usual 4D Dirac gamma matrices $\{\gamma^\mu\}$. The 5D gauge fields can be Fourier expanded about the OUED coordinate as

$$A_\mu\left(x^\nu,y\right)=\frac{1}{\sqrt{\pi R}}\left[A_\mu^{(0)}\left(x^\nu\right)+\sqrt{2}\sum_{n=1}^{\infty}A_\mu^n\left(x^\nu\right)\cos\left(\frac{ny}{R}\right)\right] \tag{14}$$

The 5D gauge fields must be even under the orbifold transformation $y\to -y$, i.e. $A^\mu\left(x^\mu,-y\right)=A^\mu\left(x^\mu,y\right)$, in order to obtain their zero-modes in the effective 4D theory.

The interactions of the weak eigenstates with the zero-mode $Z$ boson are

$$L^Z=\begin{pmatrix}\bar{q}_{3L}^{(0)} & \bar{q}_{3L}^{(1)} & \bar{u}_{3L}^{(1)}\end{pmatrix}\gamma^\mu D_L^Z\begin{pmatrix}q_{3L}^{(0)}\\ q_{3L}^{(1)}\\ u_{3L}^{(1)}\end{pmatrix}Z_\mu^{(0)}+\begin{pmatrix}\bar{u}_{3R}^{(0)} & \bar{q}_{3R}^{(1)} & \bar{u}_{3R}^{(1)}\end{pmatrix}\gamma^\mu D_R^Z\begin{pmatrix}u_{3R}^{(0)}\\ q_{3R}^{(1)}\\ u_{3R}^{(1)}\end{pmatrix}Z_\mu^{(0)} \tag{15}$$

where the diagonal matrix $D_n$, which accounts for the effects from the redefinition of the fields $\left(q_3^{(0)}, q_3^{(1)}, u_3^{(1)}\right)^T \to \left(q_3^{(0)}, q_3^{(1)}, \gamma_5 u_3^{(1)}\right)^T$, has no effect on the couplings since $D_n D_{L,R}^Z D_n = D_{L,R}^Z$. The weak eigenstate coupling matrices are defined as

$$D_L^Z \equiv \frac{g}{\cos\theta_W}\begin{pmatrix} g_{4L} & 0 & 0 \\ 0 & g_{4L} & 0 \\ 0 & 0 & g_{4R} \end{pmatrix} \quad , \quad D_R^Z \equiv \frac{g}{\cos\theta_W}\begin{pmatrix} g_{4R} & 0 & 0 \\ 0 & g_{4L} & 0 \\ 0 & 0 & g_{4R} \end{pmatrix} \tag{16}$$

where $g_{4L,R}$ are the usual SM couplings of charge $2/3$ quarks to the $Z$ boson:

$$g_{4L} = -\frac{2}{3}\sin^2\theta_W \quad , \quad g_{4R} = \frac{1}{2} - \frac{2}{3}\sin^2\theta_W \tag{17}$$

Both the left-chiral and right-chiral KK excitations $q_{3L}^{(1)}$ and $q_{3R}^{(1)}$ are $SU(2)$ doublets associated with the left-chiral zero-mode doublet $q_{3L}^{(0)}$; therefore, even $q_{3R}^{(1)}$ couples as $g_{4L}$. Analogously, the KK singlets $u_{3R}^{(1)}$ and $u_{3L}^{(1)}$ associated with the observed 4D $U(1)$ singlet $u_{3R}^{(0)}$ all couple as $g_{4R}$. In terms of the mass eigenstates, Eq. 15 becomes

$$L^Z = \begin{pmatrix} \bar{t}_L & \bar{t}_{1L}^{\bullet} & \bar{t}_{1L}^{\circ} \end{pmatrix} G^{ZL} \gamma^\mu \begin{pmatrix} t_L \\ t_{1L}^{\bullet} \\ t_{1L}^{\circ} \end{pmatrix} Z_\mu^{(0)} + \begin{pmatrix} \bar{t}_R & \bar{t}_{1R}^{\bullet} & \bar{t}_{1R}^{\circ} \end{pmatrix} G^{ZR} \gamma^\mu \begin{pmatrix} t_R \\ t_{1R}^{\bullet} \\ t_{1R}^{\circ} \end{pmatrix} Z_\mu^{(0)} \tag{18}$$

where the coupling matrices $G^{ZL}$ and $G^{ZR}$ are obtained by computing $U_L^{-1} D_L^Z U_L$ and $U_R^{-1} D_R^Z U_R$, respectively. Applying the usual 4D SM projection operators, $P_{L,R} = \frac{1 \mp \gamma_5}{2}$, Eq. 18 can be cast in the form

$$L^Z = \begin{pmatrix} \bar{t} & \bar{t}_1^{\bullet} & \bar{t}_1^{\circ} \end{pmatrix} G^{ZV} \gamma^\mu \begin{pmatrix} t \\ t_1^{\bullet} \\ t_1^{\circ} \end{pmatrix} Z_\mu^{(0)} + \begin{pmatrix} \bar{t} & \bar{t}_1^{\bullet} & \bar{t}_1^{\circ} \end{pmatrix} G^{ZA} \gamma^\mu \gamma_5 \begin{pmatrix} t \\ t_1^{\bullet} \\ t_1^{\circ} \end{pmatrix} Z_\mu^{(0)} \tag{19}$$

where $g_{4V,A} = \frac{g_{4R} \pm g_{4L}}{2}$.

Although the matrix for the couplings to the $Z_\mu^{(0)}$ is diagonal in terms of the weak eigenstates (Eq. 15), it is not proportional to the identity matrix, which can lead to non-diagonal couplings to the $Z_\mu^{(0)}$ in terms of the mass eigenstates (Eq. 19) (after multiplying through with the unitary matrices). This is another unique feature of the NUH model, which gives rise to gauge interactions that violate KK number conservation.

Presently, we clarify some of our notation. The zero-mode charge $2/3$, third-generation quark fields include a left-chiral doublet $q_{3L}^{(0)} \equiv \begin{pmatrix} u_{3L}^{(0)} \\ d_{3L}^{(0)} \end{pmatrix}$ and right-chiral singlets $u_{3R}^{(0)}$ and $d_{3R}^{(0)}$. Here, $u_3^{(0)}$ and $d_3^{(0)}$ are the top and bottom quark weak eigenstates, respectively. When we write an interaction of the form $\frac{g}{\sqrt{2}}\bar{q}_{3L}^{(0)}\gamma^\mu q_{3L}^{(0)} W_\mu^{+(0)}$ (which will follow shortly), we really mean just the charge $2/3$ component of the doublet $\bar{q}_{3L}^{(0)}$ and just the charge $-1/3$ component of the doublet $q_{3L}^{(0)}$. Associated with $q_{3L}^{(0)}$ are two towers of KK excitations, for which the lowest-lying states include a left-chiral doublet $q_{3L}^{(1)}$ and a right-chiral doublet $q_{3R}^{(1)}$ (because the 5D quark fields are vector-like fields). Associated with $u_{3R}^{(0)}$ and $d_{3R}^{(0)}$ are the right-chiral KK singlets $u_{3R}^{(1)}$ and $d_{3R}^{(1)}$ and the left-chiral singlets $u_{3L}^{(1)}$ and $d_{3L}^{(1)}$. It should be clear from the context when we use $q_{3L}^{(1)}$ or $q_{3R}^{(1)}$ to refer specifically to just the charge $2/3$ or charge $-1/3$ component of the doublet. The mass eigenstates are instead denoted by $\{t, t_1^{\bullet}, t_1^{\circ}\}$ and $\{b, b_1^{\bullet}, b_1^{\circ}\}$. We neglect any mixing in the bottom sector since the bottom mass is insignificant compared to the compactification scale, so $\{b, b_1^{\bullet}, b_1^{\circ}\} = \{q_3^{(0)}, q_3^{(1)}, d_3^{(1)}\}$. However, mixing in the top sector does affect bottom interactions.

The interactions of the charge 2/3, third-generation quark fields with the zero-mode $W^{\pm}$ bosons are

$$L^W = \frac{g}{\sqrt{2}}\left[\begin{pmatrix}\bar{q}_{3L}^{(0)} & \bar{q}_{3L}^{(1)} & \bar{u}_{3L}^{(1)}\end{pmatrix} D_n \gamma^\mu \begin{pmatrix} q_{3L}^{(0)} \\ q_{3L}^{(1)} \\ d_{3L}^{(1)} \end{pmatrix} W_\mu^{+(0)} + \begin{pmatrix}\bar{q}_{3L}^{(0)} & \bar{q}_{3L}^{(1)} & \bar{d}_{3L}^{(1)}\end{pmatrix} D_n \gamma^\mu \begin{pmatrix} q_{3L}^{(0)} \\ q_{3L}^{(1)} \\ u_{3L}^{(1)} \end{pmatrix} W_\mu^{-(0)}\right] + \text{h.c.} \tag{20}$$

Expressing Eq. 20 in terms of the mass eigenstates,

$$L^W = \begin{pmatrix}\bar{t}_L & \bar{t}_{1L}^{\bullet} & \bar{t}_{1L}^{\circ}\end{pmatrix} G^{W^+} \gamma^\mu \begin{pmatrix} b_L \\ b_L^{\bullet} \\ b_L^{\circ} \end{pmatrix} W_\mu^{+(0)} + \begin{pmatrix}\bar{b}_L & \bar{b}_{1L}^{\bullet} & \bar{b}_{1L}^{\circ}\end{pmatrix} \gamma^\mu G^{W^-} \begin{pmatrix} t_L \\ t_{1L}^{\bullet} \\ t_{1L}^{\circ} \end{pmatrix} W_\mu^{-(0)} + \text{h.c.} \tag{21}$$

Eq. 21 shows that the $W_\mu^{\pm(0)}$ can couple to a zero-mode bottom quark and a KK excitation of the top quark through mixing in the top quark sector. The elements of $G^{W^\pm}$ provide the couplings of the mass eigenstates to the $W_\mu^{\pm(0)}$, and are related to $U_L$ and $U_L^\dagger$ through the factors of $g/\sqrt{2}$ and the $D_n$ matrices in Eq. 20.

There is, however, no inter-mode mixing in the electromagnetic interactions: The couplings of the weak eigenstates to the photon are the same for each state (in contrast to the couplings to the $Z_\mu^{(0)}$ in Eq. 15). Thus, the matrix for the couplings of the mass eigenstates to the photon is diagonal (in contrast to the couplings to the $Z_\mu^{(0)}$ in Eq. 19). As a result, KK number is conserved in the interactions with the photon and KK excitations must come in pairs in the couplings to the photon. For the same reason, there is no inter-mode mixing in the couplings of quarks to gluons.

Continuing our numerical example from the previous section to include the gauge interactions – where we chose $R = (500\,\mathrm{GeV})^{-1}$, $\tilde{y}_4^{33} = 1.35$, and $R_H = 0$ – the matrices for the vector and axial couplings of the mass eigenstates to the $Z_\mu^{(0)}$ are

$$G^{ZV} = \begin{pmatrix} 0.106 & 0 & 0 \\ 0 & 0.106 & 0 \\ 0 & 0 & 0.106 \end{pmatrix} \quad , \quad G^{ZA} = \begin{pmatrix} -0.151 & -0.117 & -0.200 \\ -0.117 & -0.168 & 0.186 \\ -0.200 & 0.186 & 0.043 \end{pmatrix} \qquad (22)$$

Only the axial vector couplings include off-diagonal terms. For $R_H = 0$, the matrix for the vector couplings, $G^{ZV}$, is proportional to the identity matrix. Ultimately, the reason for this can be traced back to the fact that the mass matrix in Eq. 4 can be made Hermitian by simple row/column operations (more details are provided in the Appendix). The couplings of the mass eigenstates to the $W_\mu^{\pm(0)}$ are

$$G^{W^-} = \begin{pmatrix} 0.398 & 0.183 & 0.143 \\ -0.076 & 0.371 & -0.263 \\ 0.220 & -0.204 & -0.350 \end{pmatrix} \quad , \quad G^{W^+} = \begin{pmatrix} 0.398 & -0.076 & 0.220 \\ 0.183 & 0.371 & -0.204 \\ 0.143 & -0.263 & -0.350 \end{pmatrix} \qquad (23)$$

The nonzero couplings of the mass eigenstates $\{t, t_1^\bullet, t_1^\circ\}$ to the $Z^{(0)}$ (top) and $W^{-(0)}$ (bottom) are illustrated in Fig. 2 as a function of the compactification scale $1/R$ for $R_H = 0$ using the central allowed values of $\tilde{y}_4^{33}$. The vector couplings are proportional to the identity matrix, so the $ttZ$, $t^\bullet t^\bullet Z$, and $t^\circ t^\circ Z$ vector couplings are equal, while the off-diagonal terms vanish. However, the axial couplings include off-diagonal terms, and the diagonal terms differ. As a result, the diagonal elements of $G^Z$ have both vector and axial couplings, while the off-diagonal elements of $G^Z$ are strictly axial. The couplings of the $W^{+(0)}$ to the mass eigenstates $\{t, t_1^\bullet, t_1^\circ\}$ are found by taking the transpose of $G^{W^-}$. In addition to introducing these KK number conservation violating couplings, the mixing between the zero-mode top quark and its lowest-lying KK excitations modifies the $ttZ$ and $tbW^\pm$ couplings from their SM values.

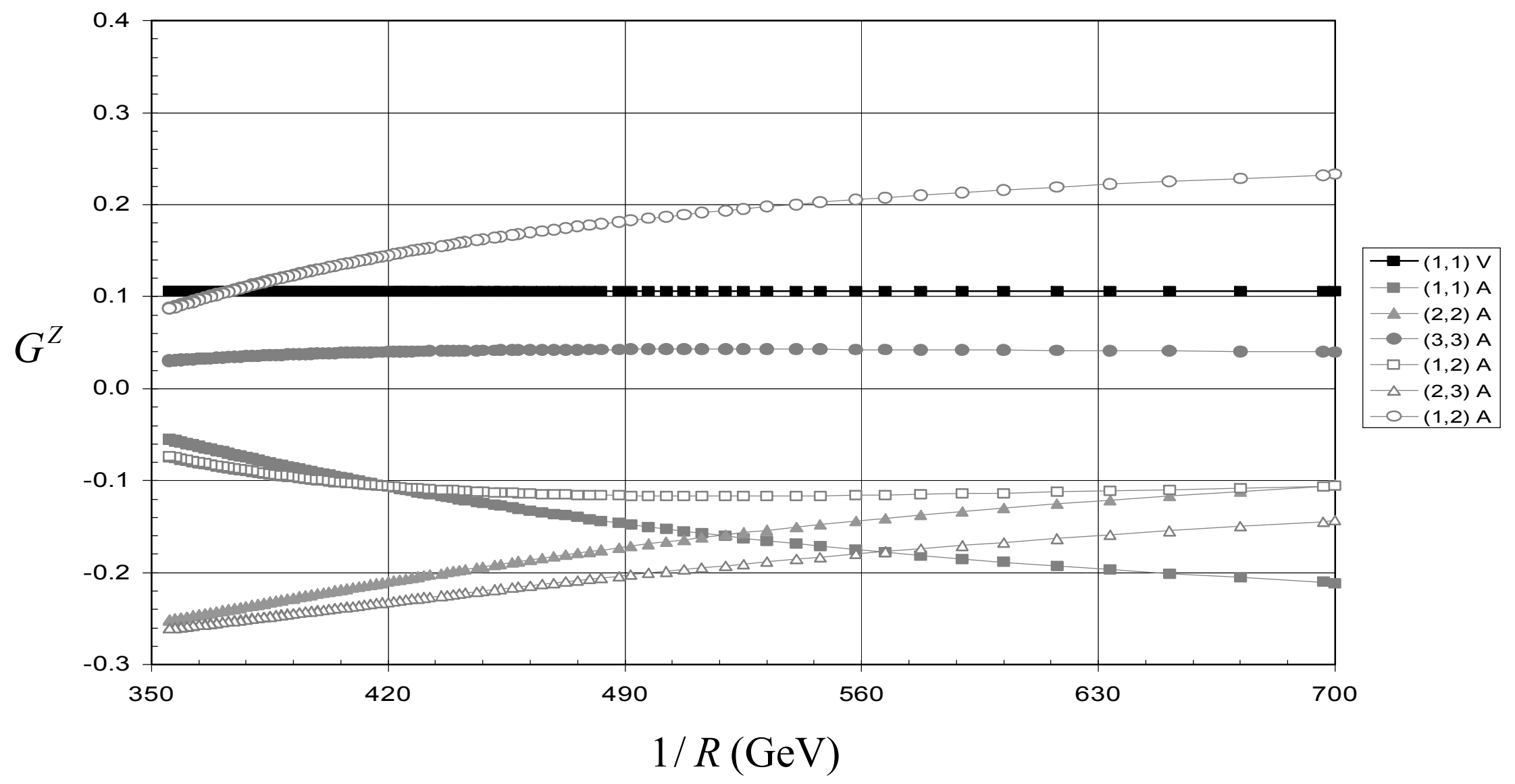

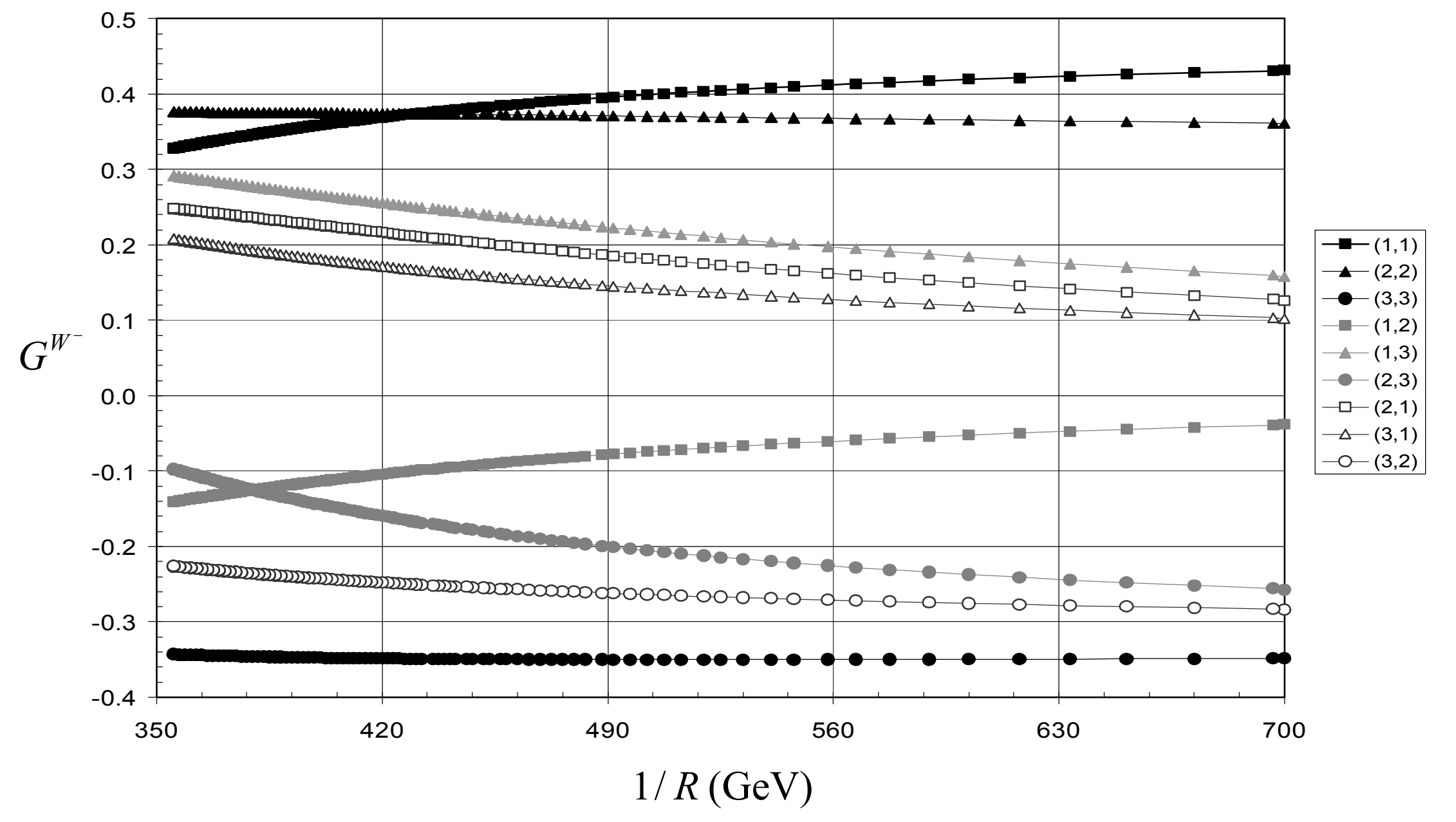

**Fig. 2**. The nonzero elements of the symmetric matrix $G^Z$ (top) and elements of the asymmetric matrix $G^{W^-}$ (bottom) for the couplings of the $t$, $t_1^\bullet$, and $t_1^\circ$ to the $Z^{(0)}$ and $W^{-(0)}$, respectively, are plotted as a function of the compactification scale $1/R$ in the NUH model with $R_H = 0$ for the central allowed values of $\tilde{y}_4^{33}$ in the basis $\left(t, t_1^\bullet, t_1^\circ\right)$. The matrix $G^{W^+}$ is the transpose of $G^{W^-}$. The branching ratios are essentially flat for $1/R$ values larger than 700 GeV.

## 4. Phenomenology

In the original UED model, KK excitations must interact in pairs in order to conserve KK number. Thus, the lightest KK excitations are stable at tree-level, which has important cosmological implications. Even the $t_1^\bullet$ (or the $t_1^\circ$, since they have degenerate mass in the original UED model) cannot decay to a $t$ and a $g_1^*$ at tree-level, for example, since $m_{t_1^\bullet} = \sqrt{m_t^2 + 1/R^2}$. The decay $t_1^\bullet \to b + W_1^*$ has high kinematic suppression, and kinematic suppression is worse for the lighter flavors. Radiative corrections [3] to the masses allow the heavier KK excitations to decay to the LKP. This leads to an abundance of LKP's in the universe.

In the NUH model, tree-level violation of KK number conservation in the Yukawa interactions – which leads to further violation in the gauge interactions – allow KK excitations of the fermions to decay to two zero-modes at tree-level via a Higgs, $Z$, or $W^\pm$. This opens up many new decay channels for the lowest-lying KK excitations. For example, some new $t_1^\bullet$ decay channels include $t_1^\bullet \to t + H$, $t_1^\bullet \to t + Z$, and $t_1^\bullet \to b + W^+$. Through mixing, the $t_1^\bullet$ can be significantly larger than $1/R$, allowing the tree-level decays $t_1^\bullet \to t + g_1^*$, $t_1^\bullet \to t + \gamma_1^*$, $t_1^\bullet \to t + Z_1^*$, and $t_1^\bullet \to b + W_1^{+*}$, which would otherwise be kinematically forbidden or highly suppressed. Also through mixing, the $t_1^\bullet$ and $t_1^\circ$ lose their degeneracy, allowing decays such as $t_1^\bullet \to t_1^\circ + Z$.[§] The principal decay modes for the $t_1^\bullet$ and $t_1^\circ$ are: $t + Z$, $b + W^+$, $t + H$, and $t + g_1^*$ (Fig. 3). The decay to a $t$ and $g_1^*$ may be kinematically forbidden for the lighter KK excitation of the top quark. The decay widths are:

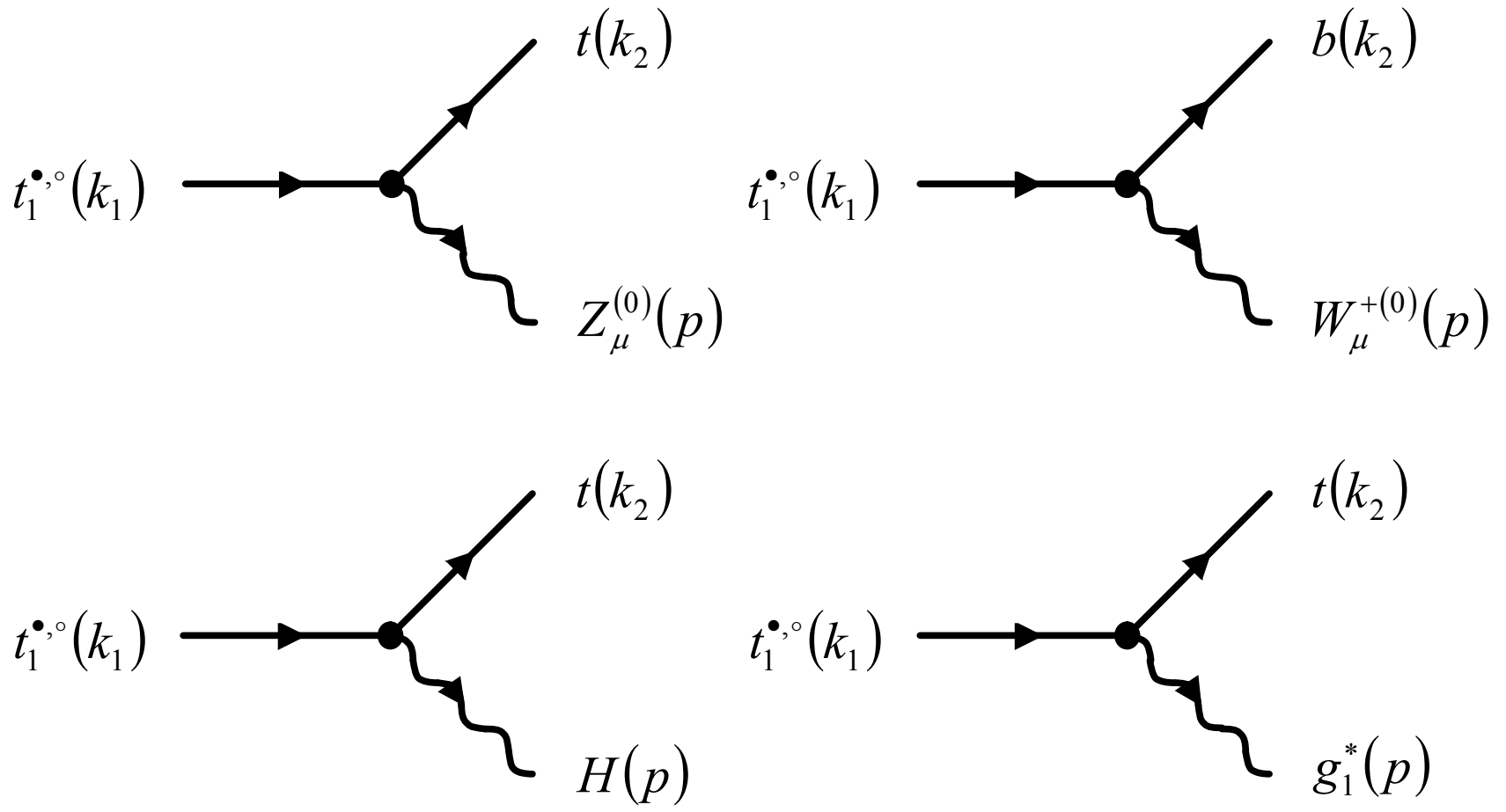

**Fig. 3**. The effective 4D tree-level Feynman diagrams for the principal decay modes of the $t_1^\bullet$ and $t_1^\circ$. One or more of the decay modes may be kinematically forbidden, especially for the lighter of the two $n = 1$ KK excitations.

[§] However, the $t_1^\bullet$ cannot decay to $t_1^\circ + g$ since the $t_1^\bullet$ - $t_1^\circ$ - $g$ coupling is zero [8].

$$
\begin{aligned}
\Gamma\left(t_1^\bullet \to t+Z\right) &= \frac{3\|\vec{\mathbf{p}}\|\left(G_{12}^{ZA}\right)^2}{4\pi m_{t_1^\bullet}}\left[\sqrt{p^2+m_t^2}+\frac{2}{m_Z^2}\sqrt{p^2+m_Z^2}\left(\sqrt{p^2+m_t^2}\sqrt{p^2+m_Z^2}+p^2\right)+3m_t\right] \\
\Gamma\left(t_1^\bullet \to b+W^+\right) &= \frac{3\|\vec{\mathbf{p}}\|\left(G_{21}^{W^+}\right)^2}{8\pi m_{t_1^\bullet}}\left[\|\vec{\mathbf{p}}\|+\frac{2}{m_W^2}\sqrt{p^2+m_W^2}\left(p\sqrt{p^2+m_W^2}+p^2\right)\right] \\
\Gamma\left(t_1^\bullet \to t+H\right) &= \frac{\|\vec{\mathbf{p}}\|}{4\pi m_{t_1^\bullet}}\left(G_{12}^{H}\right)^2\left(\sqrt{p^2+m_t^2}+m_t\right) \\
\Gamma\left(t_1^\bullet \to t+g_1^*\right) &= \frac{g_S^2\|\vec{\mathbf{p}}\|}{2\pi m_{t_1^\bullet}}\left[\sqrt{p^2+m_t^2}+\frac{2}{m_{g_1^*}^2}\sqrt{p^2+m_{g_1^*}^2}\left(\sqrt{p^2+m_t^2}\sqrt{p^2+m_{g_1^*}^2}+p^2\right)\right]
\end{aligned}
\tag{24}
$$

where we have neglected the mass of the zero-mode bottom quark. The decay widths for the $t_1^\circ$ involve the following replacements: $G_{12}^{ZA} \to G_{13}^{ZA}$, $G_{21}^{W^+} \to G_{31}^{W^+}$, and $G_{12}^{H} \to G_{13}^{H}$.

The branching ratios for the principal decay modes of the $t_1^\bullet$ (black) and $t_1^\circ$ (gray) are illustrated in Fig. 4 as a function of the compactification scale $1/R$ with $R_H = 0$ and $m_H = 200\,\text{GeV}$ for the central allowed values of $\tilde{y}_4^{33}$. In this range of parameter space, the $t_1^\circ$ – which is the lighter KK mass eigenstate – has at most three principal decay modes since it does not have enough mass to decay to $t + g_1^*$.

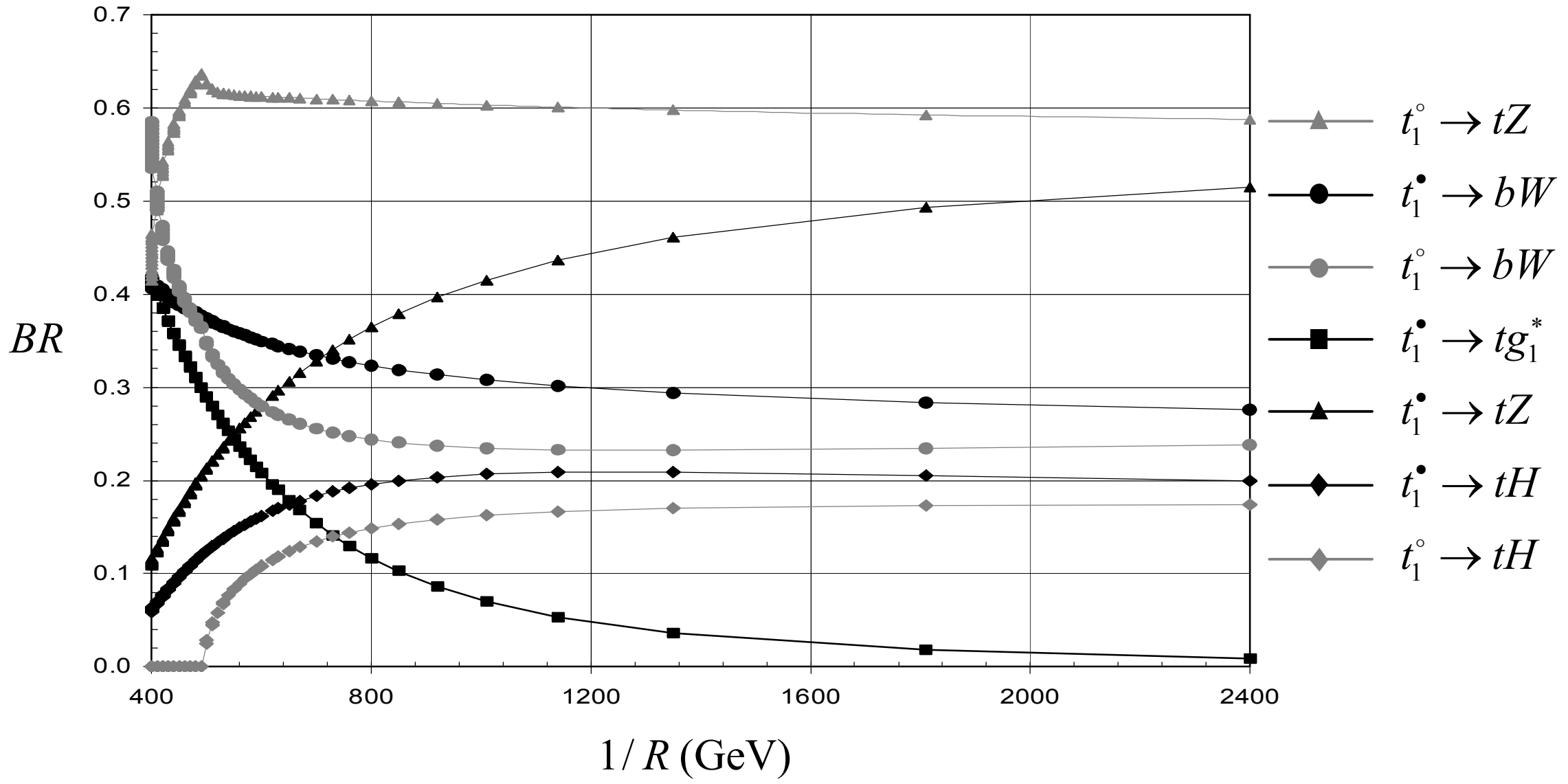

**Fig. 4**. The branching ratios for the principal decay modes of the $t_1^\bullet$ (black) and $t_1^\circ$ (gray) are plotted as a function of the compactification scale $1/R$ in the NUH model with $R_H = 0$ for the central allowed values of $\tilde{y}_4^{33}$. The $t_1^\circ$ does not have enough mass to decay to $t + g_1^*$ in this range of parameter space.

Mixing in the top sector also allows the $b_1^\bullet$ and $b_1^\circ$ to decay to two zero-modes at tree-level, even in the limit that there is no mixing in the bottom sector (even without imposing this limit, mixing in the bottom sector is negligible compared to the top sector since the observed bottom quark mass is very tiny in comparison to the compactification scale). The $b_1^\bullet$ can decay via $b_1^\bullet \to t + W^-$ with the coupling given in Eq. 21. The lowest-lying KK excitations of the lighter generations will also be able to decay to two zero-modes, for although the mixing is more prominent for the top quark, the off-diagonal terms in the Yukawa and gauge interactions will still be non-zero for the first and second generations. The decay width for $b_1^\bullet \to t + W^-$ is

$$\Gamma\left(b_1^\bullet \to t + W^-\right) = \frac{3\left|\vec{\mathbf{p}}\right|\left(G_{12}^{W^-}\right)^2}{8\pi m_{b_1^\bullet}}\left[\sqrt{p^2 + m_t^2} + \frac{2}{m_W^2}\sqrt{p^2 + m_W^2}\left(\sqrt{p^2 + m_t^2}\sqrt{p^2 + m_W^2} + p^2\right)\right] \quad (26)$$

The $b_1^\circ$ decay is the same except for the replacement $G_{12}^{W^-} \to G_{13}^{W^-}$.

Violation of KK number conservation in the NUH model has a significant effect on the collider phenomenology. One striking feature is that KK excitations can be produced singly via interactions with the Higgs or weak gauge bosons. The lowest-lying KK excitations of the top and bottom could be produced singly at the LHC by exchanging a zero-mode $Z$ – e.g. $u\bar{u} \to t_1^\bullet \bar{t}$ – or a zero-mode $W^\pm$ – e.g. $u\bar{d} \to t_1^\bullet \bar{b}$ or $d\bar{u} \to b_1^\bullet \bar{t}$. However, processes like $u\bar{u} \to b_1^\bullet \bar{b}$, which do not involve a top quark, are highly suppressed in comparison because mixing in the bottom sector is significantly smaller than mixing in the top sector (since $m_b << 1/R$); the production of KK excitations of the lighter generations is similarly suppressed. Although these weak cross sections are suppressed by a factor of $O\left(\alpha^2 / \alpha_S^2\right)$ compared to the corresponding strong cross sections due to differences in coupling strengths, the final states are lighter (since in the original UED model the KK excitations are pair-produced) and they may stand out well against the SM background because the KK excitations will often decay via weak interactions. For example, $t_1^\bullet \bar{t}$ production can result in $t\bar{b}W^-$, $b\bar{t}W^+$, $t\bar{t}Z$, or $t\bar{t}H$ via subsequent decay of the $t_1^\bullet$. The SM contribution to $t\bar{t}Z$, for example, has three vertices at tree-level, whereas the KK contribution has two vertices at tree-level (the third final state arises from a subsequent decay), which reduces the cross section only by a factor of the branching ratio for $t_1^\bullet \to t + Z$.

The production rates for two zero-modes – namely, $t\bar{t}$, $t\bar{b}$, and $b\bar{t}$ – are also affected through mixing in the gauge sector – i.e. the $(1,1)$ elements of the matrices for the gauge couplings are modified by the mixing of the third generation, charge 2/3 zero-mode quark with its KK excitations. The cross sections for $t\bar{t}$, $t\bar{b}$, and $b\bar{t}$ production at the LHC or a future $e^+e^-$ collider can be compared with the predictions of the SM and the NUH model.

The $t_1^\bullet$, $t_1^\circ$, $b_1^\bullet$, and $b_1^\circ$ would be produced most abundantly at the LHC in pairs via a virtual gluon – e.g. $u\bar{u} \to t_1^\bullet \bar{t}_1^\bullet$ or $u\bar{u} \to b_1^\bullet \bar{b}_1^\bullet$. Although the final states are heavier than when the KK excitations are produced singly, the couplings are all $O(\alpha_S)$. Although mixing is much greater in the top sector than the bottom sector, $b_1^\bullet \bar{b}_1^\bullet$ production is equally significant compared to $t_1^\bullet \bar{t}_1^\bullet$ production since the $b_1^\bullet$ can decay via $b_1^\bullet \to t + W^-$ through the top quark mixing. In the NUH model, the $t_1^\bullet$, $t_1^\circ$, $b_1^\bullet$, and $b_1^\circ$ will frequently decay via a weak or Higgs channel, which helps $t_1^\bullet \bar{t}_1^\bullet$, $t_1^\circ \bar{t}_1^\circ$, $b_1^\bullet \bar{b}_1^\bullet$, and $b_1^\circ \bar{b}_1^\circ$ production stand out well against the SM background compared to the original UED model. The cross section for $t_1^\bullet \bar{t}_1^\bullet$ and $t_1^\circ \bar{t}_1^\circ$ production can lead to $t\bar{t}ZZ$, $t\bar{t}ZH$, $t\bar{t}HH$, $b\bar{t}ZW^+$, $t\bar{b}ZW^-$, or $b\bar{b}W^+W^-$, as well as processes where the $t_1^\bullet$ or $t_1^\circ$ decay to $t + g_1^*$; and $b_1^\bullet \bar{b}_1^\bullet$ and $b_1^\circ \bar{b}_1^\circ$ production will lead to $t\bar{t}W^+W^-$.

The LHC cross sections for $b_1^\bullet \bar{b}_1^\bullet$ and $b_1^\circ \bar{b}_1^\circ$ production are a little smaller than the LHC cross section for $t_1^\circ \bar{t}_1^\circ$ production yet is somewhat larger than the LHC cross section for $t_1^\bullet \bar{t}_1^\bullet$ production because of the effect that the mixing between the zero-mode and lowest-lying KK excitations of the top quark has on the top quark masses – i.e. $m_{t_1^\circ} < m_{b_1^\bullet} \approx m_{b_1^\circ} < m_{t_1^\bullet}$ for much of the allowed region of parameter space. Since the mixing is negligible for the bottom quark mass, and since the mixing does not affect the strong couplings, the production rates for $b_1^\bullet \bar{b}_1^\bullet$ and $b_1^\circ \bar{b}_1^\circ$ are identical to the UED cross sections for the lowest-lying KK excitations of massless quark fields [8]. The significant difference is that, in the NUH model, the $b_1^\bullet$ and $b_1^\circ$ each decay predominantly to a $t$ and a $W^-$. Thus, the LHC production rates for $b_1^\bullet \bar{b}_1^\bullet$ and $b_1^\circ \bar{b}_1^\circ$ will be equal and will lead to $t\bar{t}W^+W^-$ production through their subsequent decays. Since $t\bar{t}W^+W^-$ is produced through the strong interaction and subsequent decays, it may stand out well against the SM background, which is a semi-weak process with four vertices.

The KK contributions to the LHC cross section are plotted in Fig. 5 for the various final state possibilities. Among the largest cross sections are: $t\bar{t}W^+W^-$ production (gray circles ●), corresponding to $b_1^\bullet \bar{b}_1^\bullet$ and $b_1^\circ \bar{b}_1^\circ$ production since the $b_1^\bullet$ and $b_1^\circ$ each decay predominantly to a $t$ and a $W^-$; $b\bar{b}W^+W^-$ (black diamonds ♦), $t\bar{t}ZZ$ (black triangles ▲), and $t\bar{b}ZW^- + b\bar{t}ZW^+$ production (gray rectangles ■); and, if the compactification scale exceeds ~500 GeV, $t\bar{t}HZ$ (white rectangles □) and $t\bar{b}HW^- + b\bar{t}HW^+$ production (white diamonds ◊). The cross section for $t_1^\circ \bar{t}_1^\circ$ significantly exceeds that of $t_1^\bullet \bar{t}_1^\bullet$ because the $t_1^\circ$ is significantly lighter.

We now compare the cross sections for these final states with the Standard Model backgrounds. The Standard Model contributions for many of these final states have been calculated [12]. For example, for the $t\bar{t}W^+W^-$ final states at the LHC energy, $\sigma_{SM} \sim 20\,\text{fb}$, compared to our KK signal cross sections of $\sim (70, 30, 1)\,\text{pb}$ for the values of the respective compactification scales, $1/R = (490, 560, 700)\,\text{GeV}$. For the $t\bar{t}ZZ$ final states, at the LHC energy, $\sigma_{SM} \sim 2\,\text{fb}$, compared to our KK signal cross sections of $\sim (60, 30, 0.6)\,\text{pb}$ for $1/R = (490, 560, 700)\,\text{GeV}$, respectively. Thus these KK signals will stand out very well against the SM backgrounds. Note that the reason that the SM background is so small is because the SM contributions involve pairs weak production vertices, while our signals arises from the production of the KK pairs via strong interactions, and their subsequent decays. This is true even in light of uncertainties in the computation of the LHC cross section, which may be as high as about fifty percent due to uncertainties in the CTEQ parton distribution functions [13], neglect of higher-order corrections, and the scale-dependence.

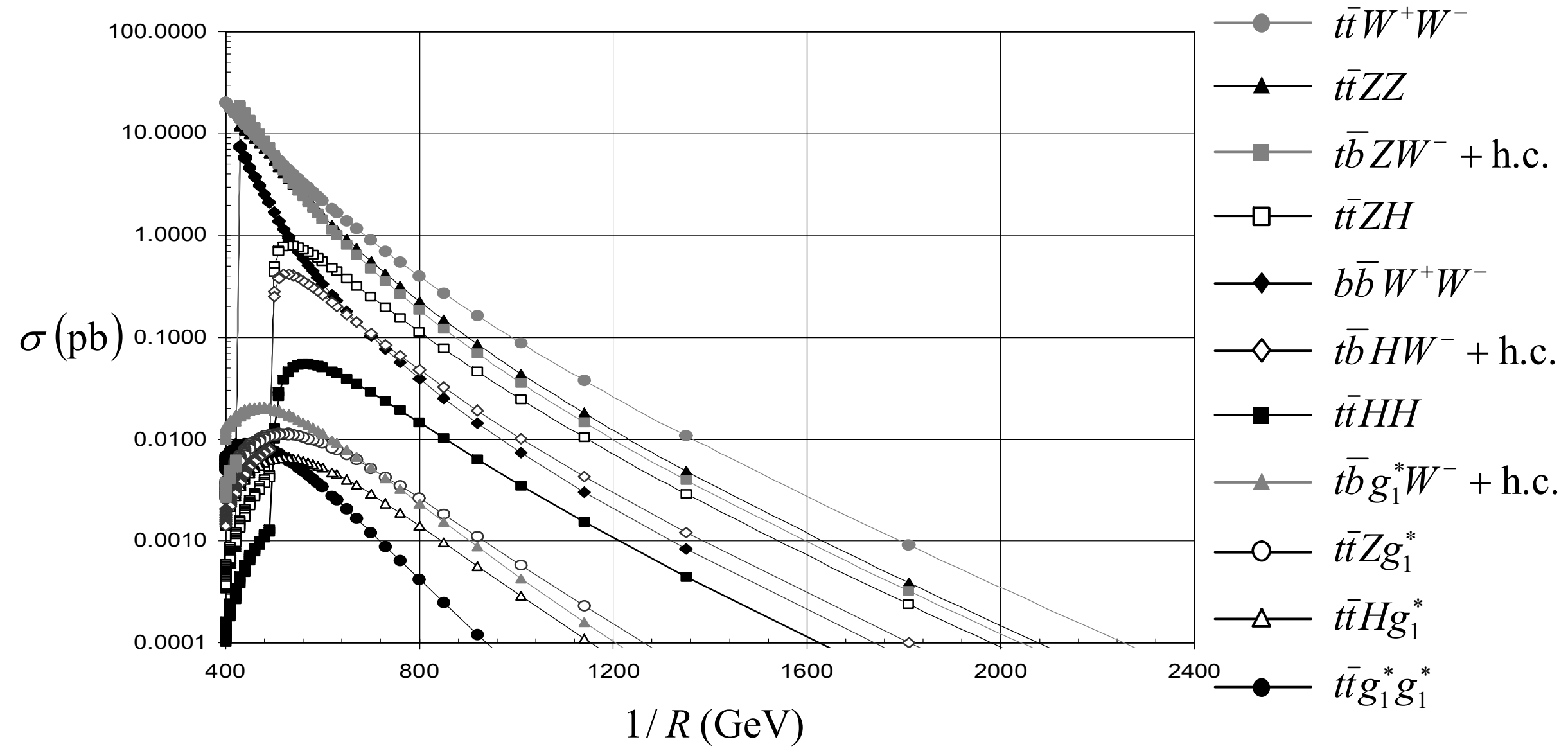

**Fig. 5**. The KK contributions to various LHC cross sections for $t_1^\bullet \bar{t}_1^\bullet$, $t_1^\circ \bar{t}_1^\circ$, $b_1^\bullet \bar{b}_1^\bullet$, and $b_1^\circ \bar{b}_1^\circ$ production are plotted as a function of the compactification scale $1/R$ in the NUH model with $R_H = 0$ for the central allowed values of $\tilde{y}_4^{33}$. The jumps at around 420 GeV correspond to where the $t_1^\circ$ does not have enough mass to decay on-shell to produce the indicated final state.

Note that even for values of $1/R$ around 2 TeV, there are observable signals for the $t\bar{t}W^+W^-$ and $t\bar{t}ZZ$ final states at the LHC energy. These final state $t$'s, $\bar{t}$'s, $W$'s and $Z$'s arise from the decays of very heavy KK modes of masses around 2 TeV. Thus, they will have very high $p_T$. These same final state particles arising from the SM will have much lower $p_T$ and hence the SM background for these final states will be negligible.

## 5. Conclusions

We have considered in detail how a non-universal Higgs with one OUED affects conservation of KK number and how this results in many striking features compared to the original UED model. In particular, we showed that KK number conservation is not only violated in the Yukawa interactions, but also in the gauge interactions. Two interesting features in the NUH model make this possible: First, the matrix for the Yukawa interactions differs from the mass matrix due to the $\pm 1/R$ contributions from the 5D kinetic terms; and second, the $q_{3L}^{(0)}$, $q_{3L}^{(1)}$, and $u_{3L}^{(1)}$ couple differently in the gauge interactions – a consequence that the 5D quark fields are vector-like four-component fermions – which places a diagonal matrix, differing from the identity, between $U_L^{-1}$ and $U_L$ in the gauge interactions (and similarly for the right-chiral fields).

We computed the mixing between the zero-mode and lowest-lying KK excitations of the third-generation, charge 2/3 quark fields, for which the mixing is dominant since the observed top quark mass is the only zero-mode quark mass which may be significant compared to the compactification scale. We used this mixing matrix to calculate the matrices for the couplings of the top quark and its lowest-lying KK excitations to the Higgs and gauge bosons. In contrast to the original UED model, this mixing – which can be quite large – allows couplings involving a single KK excitation. This leads to much different collider phenomenology for the NUH model.

We calculated the branching ratios for the most prominent decay channels of the lowest-lying KK excitations of the top and bottom quarks. We found that the $t_1^\bullet$, $t_1^\circ$, $b_1^\bullet$, and $b_1^\circ$ can decay frequently via a weak or Higgs coupling, which helps to filter the KK signal from the SM background. We considered the collider phenomenology of the NUH model. As a result of violation of KK number conservation in the gauge interactions, the KK excitations can be produced singly via a virtual $Z$ or $W^\pm$ boson (whereas they must be pair-produced at tree-level in the original UED model). Also, the cross sections for the production of two zero modes – $t\bar{t}$, $t\bar{b}$, and $b\bar{t}$, in particular – are affected through mixing in the top sector in the NUH model. These cross sections can be compared with production rates at the LHC or a future $e^+e^-$ collider to place bounds on the parameters of the NUH model. Finally, we computed the LHC cross sections for $t_1^\bullet \bar{t}_1^\bullet$, $t_1^\circ \bar{t}_1^\circ$, $b_1^\bullet \bar{b}_1^\bullet$, and $b_1^\circ \bar{b}_1^\circ$ production, which would be the most abundant production of $t_1^\bullet$, $t_1^\circ$, $b_1^\bullet$, and $b_1^\circ$ quarks. Many of the final state signal cross sections, arising from the subsequent decays of the KK pairs via the KK violating gauge interactions, stand out well against the SM background, and will be observable at the LHC for a wide range of compactification scales.

We thank Tao Han for a useful communication. We also thank G. Burdman and Z. Chacko for a very useful communication regarding the constraint on the compactification scale arising from the EW precision data for the $T$-parameter. CM would like to thank OSU for its kind hospitality during this research. SN would like to thank the members of the High Energy Theory Group of Brookhaven National Laboratory for warm hospitality and support during this summer when part of this work was carried out. The research of SN is supported in part by the US Department of Energy Grant numbers DE-FG02-04ER41306 and DE-FG02-04ER46140.

**Appendix**

The bi-unitary transformation is given by

$$D_n U_L^{-1} M_t U_R = M_D \qquad (A1)$$

where $D_n \equiv diag\{1,1,-1\}$. If $R_H = 0$, then the mass matrix $M_t$ can be made Hermitian by simple row/column operations (see Eq. 4):

$$M_t' \equiv M_t R_c(2 \leftrightarrow 3) D_n \qquad (A2)$$

where the operator $R_c(2 \leftrightarrow 3)$ swaps the 2nd and 3rd columns of a matrix when multiplying from the right of a matrix:

$$R_c(2 \leftrightarrow 3) = \begin{pmatrix} 1 & 0 & 0 \\ 0 & 0 & 1 \\ 0 & 1 & 0 \end{pmatrix} \qquad (A3)$$

When multiplying from the left of a matrix, $R_c(2 \leftrightarrow 3)$ instead interchanges the 2nd and 3rd rows.

Since $M_t'$ is a Hermitian matrix (unlike $M_t$), there exists a unitary transformation of the form

$$D_n U_L'^{-1} M_t' U_L' = M_D \qquad (A4)$$

Note that $U_R' = U_L'$ and $M_t'^{\dagger} = M_t'$ in the case of the Hermitian matrix. Thus,

$$D_n U_L'^{-1} M_t' U_R' = D_n U_L'^{-1} M_t R_c(2 \leftrightarrow 3) D_n U_R' = M_D \qquad (A5)$$

One way that Eq.'s A1 and A5 can agree is if the unprimed and primed unitary matrices are related as follows:

$$\begin{aligned} U_L' &= U_L \\ U_R' &= D_n R_c(2 \leftrightarrow 3) U_R \end{aligned} \qquad (A6)$$

Note, for example, that $D_n M_t D_n = M_t$. Thus, the unitary matrices $U_L$ and $U_R$ are related by

$$\begin{aligned} U_L &= D_n R_c(2 \leftrightarrow 3) U_R \\ U_R &= R_c(2 \leftrightarrow 3) D_n U_L \end{aligned} \qquad (A7)$$

Thus, the matrix $G_{LR}^{Y}$ can be expressed in terms of $U_L$ as:

$$\begin{aligned} G_{LR}^{Y} &= D_n U_L^{-1} Y U_R = 2 D_n U_L^{-1} M_t U_R / \upsilon - 2 D_n U_L^{-1} K R_c (2 \leftrightarrow 3) D_n U_L / \upsilon \\ G_{LR}^{Y} &= 2 M_D / \upsilon - 2 D_n U_L^{-1} K R_c (2 \leftrightarrow 3) D_n U_L / \upsilon \end{aligned} \qquad (A8)$$

using $\upsilon Y/2 = M_t - K$. Similarly,

$$\begin{aligned} G_{RL}^{Y} &= U_R^{-1} Y U_L D_n = U_R^{-1} M_t^{\dagger} U_L D_n - 2 U_L^{-1} D_n R_c (2 \leftrightarrow 3) K U_L D_n / \upsilon \\ G_{RL}^{Y} &= 2 M_D / \upsilon - 2 U_L^{-1} D_n R_c (2 \leftrightarrow 3) K U_L D_n / \upsilon = G_{LR}^{Y} \end{aligned} \qquad (A9)$$

since $K R_c (2 \leftrightarrow 3) D_n = D_n R_c (2 \leftrightarrow 3) K$. Therefore, the scalar couplings $\left(G_{LR}^{Y} + G_{RL}^{Y}\right)/2$ equal $G_{LR}^{Y}$, while the pseudoscalar couplings $\left(G_{LR}^{Y} - G_{RL}^{Y}\right)\gamma_5/2$ vanish.

A similar situation arises with the gauge couplings. The matrix $G_R^Z$ can be expressed in terms of $U_L$ as:

$$\begin{aligned} G_R^Z &= U_R^{-1} D_R^Z U_R = U_L^{-1} D_n R_c (2 \leftrightarrow 3) D_R^Z R_c (2 \leftrightarrow 3) D_n U_L \\ G_R^Z &= U_L^{-1} R_c (2 \leftrightarrow 3) D_R^Z R_c (2 \leftrightarrow 3) U_L \\ G_R^Z &= U_L^{-1} R_c (2 \leftrightarrow 3) \frac{g}{\cos\theta_W} \begin{pmatrix} g_{4R} & 0 & 0 \\ 0 & g_{4L} & 0 \\ 0 & 0 & g_{4R} \end{pmatrix} R_c (2 \leftrightarrow 3) U_L \\ G_R^Z &= U_L^{-1} \frac{g}{\cos\theta_W} \begin{pmatrix} g_{4R} & 0 & 0 \\ 0 & g_{4R} & 0 \\ 0 & 0 & g_{4L} \end{pmatrix} U_L \end{aligned} \qquad (A10)$$

The last step follows since the $R_c(2 \leftrightarrow 3)$ on the right of $D_R^Z$ interchanges its 2nd and 3rd columns, and the $R_c(2 \leftrightarrow 3)$ on the left of $D_R^Z$ interchanges its 2nd and 3rd rows: The combined effect is to swap the 2nd and 3rd diagonal elements of $D_R^Z$. The matrix $G_L^Z$ is

$$G_L^Z = U_L^{-1} \frac{g}{\cos\theta_W} \begin{pmatrix} g_{4L} & 0 & 0 \\ 0 & g_{4L} & 0 \\ 0 & 0 & g_{4R} \end{pmatrix} U_L \qquad (A11)$$

The matrix $G_V^Z$ is therefore proportional to the identity operator:

$$G_V^Z = U_L^{-1} \frac{g}{2\cos\theta_W} \begin{pmatrix} g_{4L} + g_{4R} & 0 & 0 \\ 0 & g_{4L} + g_{4R} & 0 \\ 0 & 0 & g_{4R} + g_{4L} \end{pmatrix} U_L$$

$$G_V^Z = U_L^{-1} \frac{g}{2\cos\theta_W} (g_{4L} + g_{4R}) I U_L \qquad (A12)$$

$$G_V^Z = \frac{g g_{4V}}{\cos\theta_W}$$

However, the matrix $G_A^Z$ has off-diagonal terms:

$$G_A^Z = U_L^{-1} \frac{g}{2\cos\theta_W} \begin{pmatrix} g_{4R} - g_{4L} & 0 & 0 \\ 0 & g_{4R} - g_{4L} & 0 \\ 0 & 0 & g_{4L} - g_{4R} \end{pmatrix} U_L$$

$$G_A^Z = \frac{g g_{4A}}{\cos\theta_W} U_L^{-1} \begin{pmatrix} 1 & 0 & 0 \\ 0 & 1 & 0 \\ 0 & 0 & -1 \end{pmatrix} U_L \qquad (A13)$$